\documentstyle[11pt,epsfig]{article}
\begin{document}
\title{RADII AND BINDING ENERGIES OF NUCLEI FROM\\
 A MODEL OF THE $pn$-PAIR INTERACTIONS}
\author{ G.K. Nie
\\
{\it Institute of Nuclear Physics, Tashkent, Uzbekistan}\\
galani@Uzsci.net}

\date{}
\maketitle

\begin{abstract}
New formulas to calculate the nuclear charge radii and the nuclear binding energies
have been obtained in the framework of an alpha-cluster model. The particular
features of the model are  two assumptions that the proton-neutron pair interactions
are responsible for adherence between $\alpha$-clusters and that the proton and the
neutron from a proton-neutron pair have equal nuclear potentials. These assumptions
allow one to determine the charge and Coulomb radii in dependence on the amount of
proton-neutron pairs in the nucleus. Unlike the Weizs\"{a}cker's formula to
calculate nuclear binding energies, in the model the binding energies of
alpha-clusters and excess neutrons are estimated separately. The calculated values
are in a good agreement with the experimental data.

{\it keywords}: {nuclear structure; alpha-cluster model; Coulomb energy; binding
energy; charge radius}


\end{abstract}

\section{Introduction}

The alpha-cluster model, proposed by Gamov in 1929, soon was developed by the idea
that nucleus behaves as a liquid drop. The nuclear liquid drop model was explicitly
formulated by Weizs\"{a}cker's formula in 1935 [1] for calculation of the binding
energy. Later the model had been developed by taking into account rotation
deformation modes and quantum oscillations. It was shown that the large body of data
on $\alpha$-decay is described by a simple cluster model with using a notion of
$\alpha$-core potential [2]. There are also some approaches considering some regular
forms alpha clusters might build [3]. There are several attempts to incorporate the
alpha cluster model into the shell model [3,4]. Currently the idea of
alpha-clustering is actively used in the microscopic studies of the nuclear
structure of the light nuclei [5-7] in the framework of the few-body task, as well
as in studies of the effective nucleon-nucleon interaction [8]. There has been also
some activity recently to apply an alpha-cluster model to facilitate the description
of some characteristics of nuclei like the $\alpha$-particle separation energy [9]
or nuclear radii [10] for wide range of nuclei.

It is well known that the experimental charge radii are well described as depending
on the mass number due to the formula $R = const (A)^{1/3}$ Fm. The value of $const$
varies within $0.93 \div 1.27$ from the light nuclei to the heavy ones. For the
stable nuclei with $6 \leq Z \leq 82$ the average value of $const = 0.973$ Fm
obtained from fitting the experimental values of the radii of the most abundant
isotopes [10] gives a root mean square deviation $<\Delta^2>^{1/2}=0.148\rm Fm$ of
the calculated values of $R$ from the experimental values known with an average
error $\sim 0.030$ Fm. At the same time the calculations show that the radii can be
described even with a less deviation $<\Delta^2>^{1/2}=0.078\rm Fm$ by another
formula, $R = 1.017(2Z)^{1/3}$ Fm. Therefore the number of protons, or one can say
the number of $\alpha$-clusters $N_\alpha$, can be considered as being responsible
for charge radii disregarding to the number of excess neutrons.

In this connection it is interesting to modify the classic alpha cluster model in
accordance with the  idea suggesting that  $\alpha$-clusters are the ruling elements
of the nuclear structure. This work is devoted to the task of obtaining some
phenomenological formulas to calculate the radii and the binding energies of nuclei
in dependence on the number of alpha-clusters and their inter-cluster bonds. Unlike
the Weizs\"{a}cker's formula, the binding energy of excess neutrons is to be
estimated separately.

In the $\alpha$-cluster model presented in the work a nucleus is supposed to consist
of a number of proton-neutron pairs, $pn$-pairs, that are coupled into
alpha-clusters [11]. In case of an odd value of $Z$ there is one $pn$-pair on the
surface of an alpha-clustered nucleus. The idea that the last proton and the last
neutron in the nucleus with an odd value of Z form a $pn$-pair on the surface of
alpha-cluster drop can be supported by the fact that the total spin of the nuclei
with $N = Z$ in their ground states is equal to the doubled value of the
single-particle momentum of the last nucleon. This gives definite evidence for the
existence of strong correlation between the last proton and neutron.

A particular feature of the model is an assumption that the proton and the neutron
belonging to one pair have equivalent nuclear single-particle bound state potentials
(the EPN requirement [12,13]), which looks reasonable from the point of view of the
isotopic invariance of nuclear force. As a result, this assumption brings about a
definition of the difference $\Delta E_{pn}$ of the single-particle binding energies
for the proton and the neutron of one pair as the Coulomb energy of the proton in
field of the nucleus. It  can help in estimation of the radius of the last proton
position in the nucleus. The nuclear charge radii are calculated by means of a sum
of the square radii of the core and the peripheral $\alpha$-clusters weighted by the
numbers of clusters in the core and the number of clusters on the nucleus periphery.
This calculational procedure is similar to that usually used in the framework of
single-particle potential approaches assumed in shell model [10,14].

It has been found [15] from analysis of nuclei with the equal numbers  of protons
and neutrons, $N = Z$, that the number of short-range bonds between $N_\alpha$
alpha-clusters is calculated by the formula $3(N_\alpha-2)$. A good agreement
between experimental binding energies of the nuclei with $N = Z$ and the values of
energies calculated with taking into account only the short-range bonds allows one
to suggest that that portion of the Coulomb energy which comes from the long-range
interactions is compensated by the surface tension energy. This suggests a formula
for calculation of the surface tension energy.

Thus, on the basis of this model the formulas to calculate the Coulomb energy $E^C$,
the nuclear force energy $E^{nuc}$ and the surface tension energy $E^{st}$ of the
alpha-cluster nuclear matter have been derived in a consistent way.

It is suggested that the binding energy of excess neutrons for the nuclei with $Z >
20 $ can be calculated by means of a sum of the binding  energies of excess
neutron-neutron pairs, $nn$-pairs, as $\sum_{1}^{N_{nn}} E_{nn_i}$. The experimental
values of $E_{nn_i}$ are approximated by a phenomenological formula in dependence on
the number of pairs $N_{nn}$ with using two parameters. In previous work [11] the
experimental values of $E_{nn_i}$ were used to calculate the nuclear binding
energies.

As a result, the formulas do not contain fitting parameters. All the quantities used
have their own physical meaning. They are obtained from an independent analysis of
such values as $\Delta E_{pn}$, the binding energy of the lightest nuclei or by
fitting the experimental binding energies of the excess $nn$-pairs.

To test the validity of the formulas for calculation of $E^{nuc}$, $E^C$ and
$E^{st}$ independently of the binding energy of the excess neutrons, the
$\alpha$-particle separation energy, as well as the deuteron separation energy, have
been calculated. Agreement between the calculated values and experimental ones has
been shown.

\section{Model Of $pn$-pair Interactions  For Nuclei With $N=Z$}
\subsection{Nuclear binding energy and Coulomb energy}

A picture given in Fig. 1  shows the $pn$-pair bonds in the nuclei.
\begin{figure}[th]
\centerline{\psfig{file=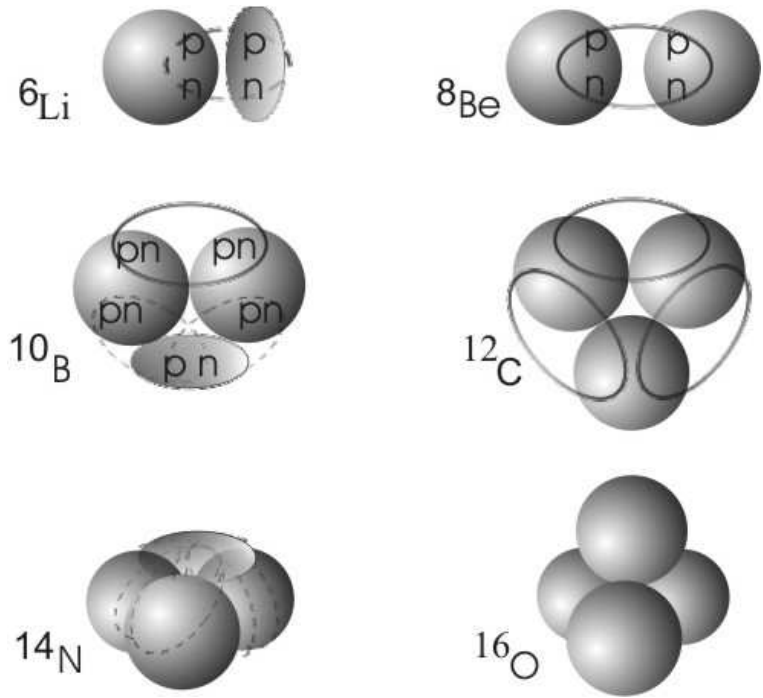,width=10cm}} \vspace*{8pt} \caption{A schematic
illustration of $pn$-pair bonds in $\alpha$-cluster model.}
\end{figure}
The internal binding energy of one cluster is taken equal to that of nucleus $^4\rm
He$ $\varepsilon_{\alpha} = 28.296\rm MeV$. The distance between clusters is
supposed to be constant for all nuclei except nucleus $^8\rm Be$. In the case one
binding is not enough to keep two positively charged clusters close. Adding one more
$pn$-pair causes two new bindings with the pairs of two alpha-clusters with the
energy $2\varepsilon_{pn pn}$ in the nucleus $^{10}\rm B$, and the clusters get
closer with the distance proper for the alpha-cluster liquid. The energy of
adherence of two clusters $\varepsilon_{\alpha\alpha}$ is estimated by means of the
following equations[11,15] for the binding energy $E^b$ of the lightest nuclei

\begin{eqnarray}
\nonumber
  E^b_{6\rm Li} &=& \varepsilon_\alpha + \varepsilon_{pn pn} + \varepsilon_{pn} \\
\nonumber
  E^b_{10\rm B} &=& 2\varepsilon_\alpha + \varepsilon_{\alpha \alpha} + 2\varepsilon_{pn pn} + \varepsilon_{pn} \\
\nonumber
  E^b_{12\rm C} &=& 3\varepsilon_{\alpha}+3\varepsilon_{\alpha \alpha}
\end{eqnarray}

So the  values of $\varepsilon_{\alpha \alpha} = 2.425 \rm MeV$, $\varepsilon_{pn
pn} = 2.037\rm MeV$, $\varepsilon_{pn} =1.659\rm MeV$, where $\varepsilon_{pn pn}$
denotes the energy between the single $pn$-pair and a $pn$-pair of the nearby
$\alpha$-cluster, $\varepsilon_{pn}$ stands for the binding energy of the proton and
the neutron in the single pair.

The binding energy of the other nuclei including $^{12}C$ are calculated with a good
accuracy by means of the following equations for the nuclei with even $Z$ with mass
number $A$
\begin{equation}
E^b = N_{\alpha}\varepsilon_\alpha + 3(N_{\alpha} - 2)\varepsilon_{\alpha \alpha},
\label{1}
\end{equation}
and for the nuclei with odd $Z_1 = Z + 1$ and the mass number $A_1 = A + 2$

\begin{equation}
E^b_1 = E^b + 6\varepsilon_{pn pn} + \varepsilon_{pn}, \label{2}
\end{equation}
where $N_\alpha$ denotes the number of alpha-clusters, $3(N_\alpha-2)$  stands for
the number of inter-cluster bonds. Although the idea that every new added cluster
causes three new bonds with three other clusters of its close vicinity was discussed
before [3], the formula to calculate the number of bonds was first obtained in [15].

Eq. (1) implies that binding energy of a nucleus $A$ is determined by the number of
clusters and the number of bonds. The energy of a nucleus $A_1$ is determined by the
energy of the nucleus $A$ plus the energy of six bonds between the single pair and
six pairs of the three clusters which get bound with that pair plus the binding
energy of the pair itself. The values of $E^b$ are presented in Table 1 in
comparison with experimental values $E_{exp}$ [16]. \vspace{1cm}
\begin{center} {Table 1. Charge radii and binding energies for nuclei with $N=Z$. Radii are given in Fm,
energies in MeV.}\end{center}
\begin{tabular}{ccccccccccccc}
\hline $Z$&$E_{exp}$&$E^b$&$\Delta E_{pn}$&$\Delta E_\alpha$&$E^C$&$E^C_W$&$E^{shr}$
&$E^{shr}_W$&$R_{exp}$&$R_{shl}$ &$R$ \\ \hline
 2& 28 &28 & .764&  .764&  .764 &  2&   29& 30 &1.71(4)$^4\rm He$&   &1.71 \\
 3& 32 &32 &1.007&      & 1.771 &  4&   34& 26 &2.57(10)$^6\rm Li$ &  &1.96 \\
 4& 56 &56 &1.644& 2.651& 3.415 &  6&   60& 61 &2.519(12)$^9\rm Be$&  &2.15 \\
 5& 65 &65 &1.850&      & 5.303 &  8&   70& 68 &2.45(12) $^{10}\rm B$& &2.32\\
 6& 92 &92 &2.764& 4.614& 8.067 & 11&  100&100 &2.470(15)$^{12}\rm C$& &2.47\\
 7&105 &106&3.003&      & 11.070& 14&  116&112 &2.560(20)$^{14} \rm N$& &2.60\\
 8&128 &128&3.536& 6.539& 14.606& 18&  142&143 &2.730(25)$^{16}\rm O$&\bf{2.73}&2.71\\
 9&137&142 &3.544&      & 18.150& 22&  156&157 &2.85$^{19}\rm F$     & 2.85 &2.82\\
10&161 &163&4.021& 7.565& 22.171& 26&  183&187 &2.992(8)$^{20}\rm Ne$ &\bf{2.91}&2.92\\
11&174&177 &4.328&      & 26.499& 31&  201&203 &2.94(4)$^{23}\rm Na$  & 2.97&2.96\\
12&198 &199&4.838& 9.166& 31.337& 35&  230&233 &3.032(28)$^{24}\rm Mg$& 3.02&3.01 \\
13&212&212 &5.042&      & 36.379& 41&  248&250 &3.06(9)$^{27}\rm Al$  & 3.07&3.05\\
14&237&234 &5.593&10.635& 41.972& 46&  279&279 &3.138(30)$^{28}\rm Si$& 3.12&3.12 \\
15&251 &248&5.731&      & 47.703& 51&  299&298 &3.24$^{31}\rm P$& 3.16&3.16 \\
16&272&270 &6.227&12.958& 53.930& 57&  326&327 &3.240(11)$^{32}\rm S$& 3.22&3.25  \\
17&285&284 &6.351&      & 60.281& 63&  346&346 &3.335(18)$^{35}\rm Cl$& 3.26&3.29 \\
18&307&306 &6.746&13.097& 67.027& 70&  374&375 &3.39$^{36}\rm Ar$& 3.36&3.38  \\
19&321&320 &6.923&      & 73.950& 76&  395&395 &3.408(27)$^{39}\rm K$& 3.39&3.41  \\
20&342 &341&7.286&14.209& 81.236& 83&  423&424 &3.482(25)$^{40}\rm Ca$&\bf{3.48}&3.51\\
21&355&355 &7.277&      & 88.513& 90&  443&444 &3.550(5)$^{45}\rm Sc$& 3.51&3.54\\
22&376& 378&7.615&14.892& 96.128& 97&  472&473 &3.59(4)$^{48}\rm Ti$& 3.55&3.64 \\
23&390& 391&7.6  &       & 103.8&104&  494&494 &3.58(4)$^{51}\rm V$& 3.58&3.67  \\
24&412& 412&7.9  & 15.5  & 111.6&112&  523&522 &3.645(5)$^{52}\rm Cr$ & 3.64&3.62 \\
25&427& 426&7.9  &       & 119.5&120&  546&544 &3.680(11)$^{55}\rm Mn$& 3.67&3.65 \\
26&448& 448&8.3  & 16.2  & 127.8&128&  575&572 &3.737(10)$^{56}\rm Fe$& 3.74&3.73 \\
27&463& 462&8.3  &       & 136.2&136&  599&594 &3.77(7)$^{59}\rm Co$ & 3.78&3.77 \\
28&484& 483&8.8  & 17.1  & 144.9&145&  629&623 &3.760(10)$^{58}\rm Ni$& 3.86&3.84 \\
29&497& 497&8.8  &       & 153.7&154&  651&645 &3.888(5)$^{63}\rm Cu$ & 3.89&3.87 \\
\hline
\end{tabular}
\vspace{1cm}

Hereafter in Tables the values are given for both even $Z$ and odd $Z_1$ nuclei with
denoting the charge and mass number as $Z$ and $A$. The values of energies and radii
in Tables and formulas are given in MeV and Fm correspondingly.

A consideration of $pn$-pairs  together with the EPN requirement gives one an
opportunity to estimate the total Coulomb energy of  nucleus $E^C$ as a sum of
differences $\Delta E_{pn}$ in the binding energies of protons and neutrons of the
$pn$-pairs

\begin{equation}
E^C = \sum^{N_{\alpha}}_1\Delta E_{\alpha} + \delta, \label{3}
\end{equation}
where $\Delta E_{\alpha}$ denotes the Coulomb energy of an $\alpha$-cluster, which
is calculated as a sum of Coulomb energies of its two $pn$-pairs $\Delta E_{\alpha}
= \sum^2\Delta E_{pn}$, $\delta$ stands for some portion of the Coulomb energy added
to the sum after the first two clusters have got closer with adding  one $pn$-pair
to them in the nucleus $^{10}\rm B$.  For the nucleus $A_1$ the corresponding
equation is

\begin{equation}
E^C_{1} = \sum^{N_{\alpha}}_1\Delta E_{\alpha} + \delta + \Delta E_{pn}. \label{4}
\end{equation}

The values of $\Delta E_{pn}$ for nuclei with $N=Z$ are calculated from experimental
binding energies taken mostly from [16], where the values are given with an accuracy
of 1 KeV with the measurement energy error in 2-3 KeV. Some of the values of mass
deficiency for light nuclei have been used from more recent tables [17]. For the
nuclei with $Z \geq 23$ the values of $\Delta E_{pn}$ cannot be obtained by this
procedure due to the lack of the experimental data for nuclei (Z, N-1). The values
were estimated [15] in framework of this model by means of the value of radius
$R=1.008(2Z)^{1/3} \rm Fm$ for the nuclei with even $Z \geq 22$. The values $\Delta
E_{pn}$, $\Delta E_{\alpha}$ and $E^C$ are given in Table 1.

The value of  Coulomb repulsion energy between two clusters $\varepsilon^C_{\alpha
\alpha} = 1.925\rm MeV$ is found from  two equations for the Coulomb energy of one
cluster in nuclei $^{12}\rm C$ and $^{16}\rm O$. They are $\Delta E_\alpha =
\varepsilon ^C_\alpha + 2\varepsilon ^C_{\alpha \alpha}$  for  $^{12}\rm C$ and
$\Delta E_\alpha = \varepsilon ^C_\alpha + 3\varepsilon ^C_{\alpha \alpha}$ for
$^{16}\rm O$, where $\varepsilon^C_\alpha$ denotes the internal Coulomb energy of
one cluster $\varepsilon^ C_\alpha = E^C_{4\rm He} = 0.764\rm MeV$.  The value of
the energy of the Coulomb repulsion between the single $pn$-pair and one of the
nearby alpha-clusters in case of odd $Z_1$ can be found from the data for the nuclei
$^6\rm Li$ and $^{14}\rm N$, $\varepsilon ^C_{pn \alpha} = 1.001(6)\rm MeV$.

Having obtained the values of the Coulomb energy of the nuclei one can estimate the
Coulomb repulsion energy between two alpha clusters in nucleus $^8\rm Be$ as follows
$\varepsilon ^C_{\alpha \alpha}(^8\rm Be) = \ E^C_{^8\rm Be} - 2\varepsilon
^C_\alpha = 1.887\rm MeV$, which  is less than $\varepsilon ^C_{\alpha \alpha}$ by
the value of $\delta = 0.038\rm MeV$.

Eqs. (3) and (4)  give the empirical values of the total Coulomb energy $E^C$ of a
nucleus obtained on the basis of the values known from the experimental data with an
accuracy of a few KeV. Having obtained the total Coulomb energy of nuclei, one can
easily test the validity of the hypothesis of the alpha-cluster structure on the
nuclei with few clusters $N_\alpha \leq 4$, that is, the nuclei where each cluster
interacts with every one of the other clusters. In this case the Coulomb energy can
be calculated by the number of clusters and their bonds. For the nucleus $^8\rm Be$
according to the Fig. 1  $E^C(^8\rm Be)= 2\varepsilon ^C_\alpha + \varepsilon
^C_{\alpha \alpha}(^8\rm Be)$. For nuclei $^{12}\rm C$ and $^{16}\rm O$ values $E^C$
are $E^C(^{12}\rm C)= 3\varepsilon ^C_\alpha + 3\varepsilon ^C_{\alpha \alpha}$ and
$E^C(^{16}{\rm O})= 4\varepsilon ^C_\alpha + 6\varepsilon ^C_{\alpha \alpha}$. In
case of the nucleus $^6\rm Li$ $E^C(^6{\rm Li}) =\varepsilon ^C_\alpha + \varepsilon
^C_{pn \alpha}$ and for $^{10}\rm B$ $E^C(^{10}{\rm B}) = 2\varepsilon^C_\alpha +
\varepsilon ^C_{\alpha \alpha} + \Delta E_{pn} $. For the nucleus $^{14}\rm N$ it is
to be $E^C(^{14}{\rm N}) = E^C(^{12}{\rm C}) + 3\varepsilon ^C_{pn \alpha}$. For
these nuclei the equations give values of $E^C$ in agreement with the values
obtained by (3) and (4) (see Table 1.) with a curious accuracy of 1 KeV with one
exception for nucleus $^6\rm Li$ (in the case the accuracy of the equation is of 6
KeV).  It confirms the validity of the cluster structure of the nuclei, as well as
the equations (3) and (4) are confirmed as the physically reliable formulas for
obtaining empirical values of the Coulomb energy with an accuracy  of several KeV.

The empirical value of the short-range nuclear force energy, which comes from
inter-cluster nuclear force bonds and the effect of surface tension, $E^{shr}=
E^{nuc} + E^{st}$ for the nuclei with $N=Z$ can be obtained from
\begin{equation}
E_{exp} = E^{shr} - E^C.\label{5}
\end{equation}

Correspondingly, the internal nuclear force energy  of one alpha-cluster is
estimated [3] as $\varepsilon ^{nuc}_\alpha = \varepsilon _\alpha + \varepsilon
^C_\alpha = 29.060\rm MeV$. In the same way the value of the nuclear force energy
between 2 alpha-clusters can be estimated as $\varepsilon ^{nuc}_{\alpha \alpha} =
\varepsilon _{\alpha \alpha} + \varepsilon ^C_{\alpha \alpha} = 2.425 \rm MeV +
1.925 MeV = 4.350\rm MeV $. The energy of nuclear force interactions between the
single $pn$-pair and the six  pairs of three nearby clusters is
$\varepsilon^{nuc}_{pn3\alpha} = 3\varepsilon ^C_{pn \alpha} + 6\varepsilon _{pn pn}
= 15.225\rm MeV$. The binding energy coming from the attracting nuclear force and
repulsing Coulomb force interaction between the single pair and three nearby
clusters is $\varepsilon_{pn3\alpha} = \varepsilon^{nuc}_{pn3\alpha} -
3\varepsilon^C_{pn3\alpha} = 12.222$ MeV.

To compare the obtained values with the values given by Weizs\"{a}cker's formula
[19]

\begin{equation}
E^b_W = \alpha A - \beta A^{2/3} + \frac{\gamma Z^2}{A^{1/3}} \pm
\frac{\delta}{A^{3/4}} - \frac{\epsilon(A/2 -Z)^2}{A}\label{6}
\end{equation}
the values of $E^{shr}_W = \alpha A - \beta A^{2/3}\pm \frac{\delta}{A^{3/4}}-
\frac{\epsilon(A/2 -Z)^2}{A}$ and $E^C_W =\frac{\gamma Z^2}{A^{1/3}}$ are given in
Table 1 in comparison with $E^{shr}$ and $E^C$. One can see that the values due to
Weizs\"{a}cker's formula and the empirical values obtained in the approach are
generally in a good agreement up to difference in a few MeV.

\subsection{Radius of last proton position  and charge radii of nuclei}

To estimate the value of the  radius of last proton position $R_p$ in nucleus $A$
the value of its Coulomb energy in the field of the nucleus is used and it is
expected to be equal to the value of $\Delta E_{pn}$. The Coulomb energy of the
proton consists of the energy of its interaction with the other proton of its
$\alpha$-cluster $\varepsilon ^C_\alpha$ and the Coulomb  energy of its interaction
with the other $Z - 2$ protons of the nucleus

\begin{equation}
\Delta E_{pn} = \varepsilon ^C_\alpha +(Z-2)e^2/R_p, \label{7}
\end{equation}
and for odd $Z_1 = Z + 1$

\begin{equation}
\Delta E_{pn} = (\varepsilon^C_{pn \alpha} - \varepsilon ^C_{\alpha}) + (Z-1)e^2/
R_{p1}, \label{8}
\end{equation}
where $R_{p1}$ stands for the radius of the position of the single $pn$-pair.

It is valid only for the nuclei with $Z/Z_1 \geq 9$, because  the last proton
Coulomb energy for the lighter nuclei must be calculated with a different function,
for example spherical function with the Coulomb radius $R_C$ like the proton Coulomb
potential [18] at small radii less than $R_C$ in the Shr\"{o}dinger equation for
single-particle bound state wave function in Distorted Wave Born Approximation.

Another way to estimate $R_p$ comes from an equation for the Coulomb energy for the
last $\alpha$-cluster. The Coulomb energy of the cluster consists of the internal
Coulomb energy of the cluster $\varepsilon ^C_\alpha$, Coulomb energy of its
interaction with the three nearby clusters $3\varepsilon ^C_{\alpha \alpha}$ and
long-range part of the Coulomb energy of its interaction with the other $N_\alpha -
4$ clusters of the nucleus

\begin{equation}
\Delta E_\alpha = \varepsilon ^C_\alpha  + 3\varepsilon ^C_{\alpha \alpha}  +
2(Z-8)e^2/ R_{N\alpha-4}, \label{9}
\end{equation}
where $R_{N\alpha-4}$ stands for the distance between the mass center of the remote
$N_\alpha - 4$ clusters and  the cluster under consideration. The value
$R_{N\alpha-4}$ is approximated as $R_{N\alpha-4} = 1.2R_p$ by fitting the empirical
values of $\Delta E_\alpha$  for the nuclei with $Z \geq 14$. For the nuclei with
odd $Z_1$ in accordance with the same logic of taking into account the long range
Coulomb interaction one gets

\begin{equation}
\Delta E_{pn} = 3\varepsilon ^C_{pn\alpha}  +  (Z-7)e^2/ (1.2R_p+R_{p1}-R_p).
\label{10}
\end{equation}

The third way to estimate $R_p$ comes from following idea. That fact that binding
energy is calculated by (1) with a good accuracy means that the long range Coulomb
energy, which increases with $Z$, must be compensated by the surface tension energy
$E^{st}_\alpha$. The latter is expected to be proportional to $R_p^2$. Therefore the
last member in the sum (9) is taken equal to $E^{st}_\alpha = \gamma_1 R_p^2 =
2(Z-8)e^2/ (1.2 R_p)$, and the value of $\gamma_1 = 0.471\ MeV/Fm^2$ is obtained
from fitting the empirical values of $\Delta E_\alpha$ for the nuclei with $Z \geq
16$. As a result one obtains an equation to calculate $R_p$ in dependence on
$N_\alpha$ only.

\begin{equation}
R_p=2.168 (N_\alpha-4)^{1/3}.  \label{11}
\end{equation}
The value of $R_{p1}$ is calculated by the same formula with the value of $N_{\alpha
1}= N_\alpha +0.5$

Finally, the fourth way to calculate $R_p$ comes from the requirement that the
surface tension energy is expected to be proportional to the number of clusters on
the surface of the liquid drop  $E^{st}_\alpha= \gamma_1 R_p^2 = \gamma_2 N_\alpha
^{2/3}$, where $\gamma_2 =1.645$ MeV is obtained by fitting the values of $\Delta
E_\alpha$ for the nuclei with $Z \geq 14$. Further for the larger nuclei with $Z
\geq 30$ the values are taken as Coulomb radius $R_C$. The equation therefore is

\begin{equation}
R_p=R_C=1.869 (N_\alpha)^{1/3}.  \label{12}
\end{equation}
For odd $Z_1$ the value of  $R_{p1} = R_{C1} $ is calculated by (12) with $N_{\alpha
1}=N_\alpha + 0.5$. The values of $R_{p/p1}$ of all four ways are equal within an
average deviation of 0.1 Fm.

The calculation of the charge radii of the nuclei has been made by two different
ways, with using shell model representation $R_{shl}$ [11,20] and by using an
representation for the core  $R$. In the shell model representation the values
$R_{shl}$ are calculated with the following equation

\begin{equation}
Z R_{shl}^2 = (Z'R_{A'}^2 + nR_p^2), \label{13}
\end{equation}
and for odd $Z_1= Z + 1$

\begin{equation}
Z_1 R_{shl1}^2 = (Z_1-1)R_A^2 +  R_{p1}^2, \label{14}
\end{equation}
where  $R_{A'}$  stands for the radius of the nucleus with completed shell $A'$ with
an atomic number $Z'$, $n$ stands for the occupation numbers for protons of the last
uncompleted shell. $A'$ denotes the nuclei $^{16}\rm O$, $^{20}\rm Ne$, $^{40}\rm
Ca$. In order to start calculation of $R_{shl/shl1}$ the value of  radius $R_{16\rm
O}$ is taken to be equal to its experimental value.  The values of $R_{shl}$ have
been calculated with the values of $R_p$ (7) and $R_{p1}$ (8).

The other way with using a representation for core is somewhat simpler, but the
deviation of calculated values from the experimental ones is even less than for
$R_{shl}$. For the light nuclei $1 \leq N_{\alpha} \leq 5$ the value $R = R_{4\rm
He} N_{\alpha}^{1/3}$, where $R_{\rm ^4 He} =1.71$ Fm [21] stands for the
experimental radius $R_{exp}$ of the nucleus $^4\rm He$. In the case of a nucleus
with odd $Z_1$ the number of clusters in the formula is  taken to be $N_{\alpha1} =
N_\alpha +0.5$.

For the nuclei with $5 < N_{\alpha} \leq 10$ the root mean square radius  $R$ is
calculated by following equation
\begin{equation}
N_{\alpha} R^2 = 5 (R_{20\rm Ne})^2 + (N_{\alpha} - 5) R_p ^2,\label{15}
\end{equation}
where $R_{p/p1}$ is calculated by the number of $N_{\alpha}/N_{\alpha 1}$ (12).

For the other nuclei with $N_{\alpha} \geq 11$ the nucleus is represented as a core
with radius $R_{\alpha} (N_{\alpha }-4)^{1/3}\rm Fm$, where $R_{\alpha} = 1.60\rm
Fm$ stands for the radius of $\alpha$-cluster of the core, plus four peripheral
$\alpha$-clusters. Then charge radius $R$ is estimated by a sum of the square radius
of the core weighted by the number of clusters in it, $N_{\alpha }-4$, and square
radius $R_p$ of the position of the last cluster (11) weighted by 4

\begin{equation}
N_{\alpha } R^2 = (N_{\alpha }-4) 1.60^2 (N_{\alpha }-4)^{2/3} + 4 R_{p}^2.
\label{16}
\end{equation}
For  odd $Z_1 = Z + 1$ the charge radius $R_1$ is calculated by equation

\begin{equation}
N_{\alpha 1} R_1^2 = N_{\alpha } R^2 + 0.5 R_{p1}^2, \label{17}
\end{equation}
where the radius of the position of the single $pn$-pair $R_{p1}$ (11) is weighted
by factor 0.5. Thus, the values of the radii calculated by the formulas  depend only
on the number of $\alpha$-clusters.

For the nuclei with $Z \leq 29$ the values of $R_{shl}$ and $R$ are given in Table 1
in comparison with $R_{exp}$ [10,21,22] for the most abundant isotopes. The values
of $R$ calculated for three nuclei $^6\rm Li$, $^9\rm Be$ and $^{10}\rm B$ are seen
not to be in agreement with the experimental data. The nuclei do not have enough
number of bonds between the clusters and the single $pn$-pair to provide a
sufficient nuclear density proper for the $alpha$-cluster liquid.

\section{Model Of $pn$-pair Interactions  For Nuclei With $Z \geq 30$}
\subsection{Charge radii}

The radii $R_{shl}$ [11] are calculated  from (13) and (14) with the values of
$R_{p/p1}$ (11). Besides the nuclei  $_{8}\rm O$, $_{10}\rm Ne$ and $_{20}\rm Ca$
the nuclei placed in the right end of the Periodical Tables Of Elements $_{28}\rm
Ni$, $_{36}\rm Kr$, $_{46}\rm Pd$, $_{54}\rm Xe$,$_{78}\rm Pt$, $_{86}\rm Rn$,
$_{110}\rm E-Pt$, $_{118}\rm E-Rn$ are taken [11] as the nuclei with the completed
shells. The nuclei $_{62}\rm Sm$, $_{70}\rm Yb$, $_{94}\rm Pu$, $_{102}\rm No$ were
added [20] to the set to make a sense that four clusters make a completed shell. For
the nuclei with the completed shells $ 28 \leq Z'\leq 70$ the value of the radius
$R_{Z'} = 1.60(N_{\alpha}')^{1/3}$, where $N_{\alpha}'$ denotes the number of
clusters in nucleus with completed shells, and for the nuclei with $Z'\geq 78$ the
value of $R_{Z'}$ is calculated from the equation $Z'R_{Z'}^2
=(Z'-2)R_{Z'-2}^2+2R_p^2$. The latter means that the fourth cluster is located above
the other three clusters completing the molecule $^{16}\rm O$, see Fig. 1.

The root mean square deviation of the calculated radii $R_{shl}$ from the
experimental values for nuclei with $9 \leq Z \leq 82$ is found to be
$<\Delta^2>^{1/2}$ =0.051 Fm.

The second way to calculate  the charge radii $R$ is to make use of (16) and (17)
with $R_{p/p1}$ (11). The obtained values of $R_{shl}$ and $R$ are presented in Fig.
2 as solid and dashed lines, correspondingly.
\begin{figure}[th]
\centerline{\psfig{file=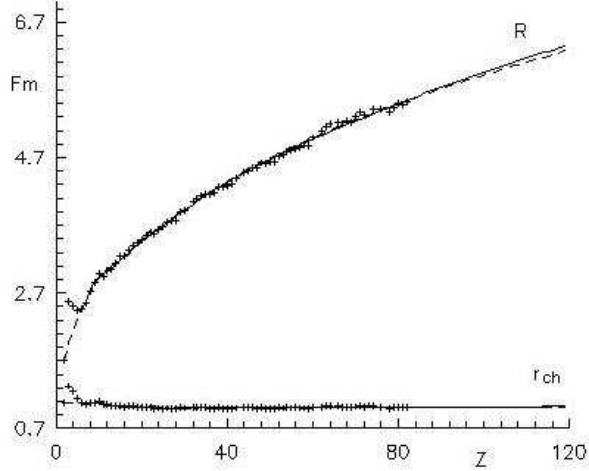,width=10cm}} \vspace*{2pt} \caption{Charge radii
$R_{shl}$ and $R$ in comparison with $R_{exp}$ known for the most abundant isotopes
and average charge radius of one nucleon $r_{ch}$ of $\alpha$ clusters in dependence
on $Z$}
\end{figure}

The experimental values $R_{exp}$ [10,21,22] for nuclei with $2 \leq Z \leq 82$ are
indicated with  crosses.  For the nuclei with $6 \leq Z \leq 83$ the deviation of
values of $R$ from experimental values $R_{exp}$   $<\Delta ^2>^{1/2} = 0.050$ Fm.
The deviation between $R_{shl}$ and $R$ for the nuclei with $9 \leq Z \leq 118$ is
$<\Delta ^2>^{1/2} = 0.034$ Fm. In Fig. 2 the values of charge radius of one nucleon
of an $\alpha$-cluster $r_{ch} = R_{shl/ /exp} /(2Z)^{1/3}\approx 1.01$ Fm are also
presented with the indications corresponding those of $R_{shl}$, $R$ and $R_{exp}$.
Fig. 2 shows that the values of $R_{shl}$ and $R$ are close to the values of
$R_{exp}$. That fact that $r_{ch}\approx$ constant approves  that the  size of a
nucleus is determined by the number of $\alpha$ clusters.

\subsection{Binding energy}
One can write Eqs. (1) and (2) in a form with revealing the  Coulomb energy and the
surface tension energy. For an alpha-cluster nucleus $A$ the binding energy consists
of the nuclear force energy $E^{nuc}$, the surface tension energy $E^{st}$, the
binding energy of excess neutrons $E_n$ and the Coulomb energy $E^C$

\begin{equation}
E^b = E^{nuc}_A  + E^{st}  + E_n - E^C, \label{18}
\end{equation}
and for nucleus  $A_1$

\begin{equation}
E^b_1 = (E^{nuc}_A + \varepsilon ^{nuc}_{pn 3\alpha} + \varepsilon_{pn}) + E^{st}_1
  + E_{n1} - E^C_1,\label{19}
\end{equation}
where $E^{nuc}_A = N_\alpha \varepsilon^{nuc}_\alpha +
3(N_\alpha-2)\varepsilon^{nuc}_{\alpha \alpha}$ is calculated by the number of
$\alpha$-clusters  and the short-range bonds  between them;

\begin{equation}
E^{st} = \sum^{N_\alpha}_{5} E^{st}_\alpha = \sum^{14}_{5} E^{st}_\alpha +
\sum^{N_\alpha -1}_{15} \gamma_1 R_C^2 + \gamma_1 R_p^2,\label{20}
\end{equation}
where the empirical values of $E^{st}_\alpha = 2(Z-8)e^2/ R_{N\alpha-4}= \Delta
E_\alpha - \varepsilon ^C_\alpha  - 3\varepsilon ^C_{\alpha \alpha}$ (9)  are
obtained for $N_\alpha \leq 14$.

In case of the nucleus with $Z_1$
\begin{equation}
E^{st}_1 = E^{st}+  \gamma_1 R_{p1}^2/2,  \label{21}
\end{equation}
where $R_{p1}$ is obtained from (11) with $N_{\alpha 1} = N_\alpha +0.5$.

The empirical values of $E^C$ and $E_1^{C}$ for nuclei with $Z \leq 29$ have been
obtained, see Table 1. For other nuclei the formula to calculate the  Coulomb energy
of a charge sphere with radius $R_C$ (12) is used

\begin{equation}
E^C  = \frac{3}{5}\frac{Z^2e^2}{R_{C}}. \label{22}
\end{equation}
For $E^C_1$ the value $R_{C1}$ (12) with $N_{\alpha 1} = N_\alpha + 0.5$ is used.

The value of  $E_n$ is estimated as $E_{n} = \sum_{i=1}^{N_{nn}}E_{nni}$, where
$N_{nn}=N_n/2$ stands for the number of the excess $nn$-pairs and $N_n$ denotes the
number of excess neutrons. The values of $E_{n1}$ are calculated similarly. The
values of $E_{nni}$ is fitted by an equation

\begin{equation}
E_{nni} = 22.5\rm MeV - 1.358 N_{nn}^{2/3}. \label{23}
\end{equation}
It is well known that the experimental values of the separation energy of $nn$-pairs
have deviations within 2-5 Mev determined by  what nucleus loses the pair, which
certainly refers to the shell effects. Despite the relatively big deviation of the
values, the empirical values of all excess $nn$-pairs energy $E_{n(exp)}$, known
from the equation
\begin{equation}
E_{n(exp)}  = E_{exp}(Z, N + N_n) - E_{exp}(Z,N) \label{24}
\end{equation}
only  for the nuclei with $N = Z$, are restricted in relatively narrow corridor. For
example, for the nuclei with $21 \leq Z \leq 29$ the value of separation energy of
two excess neutrons varies to within $E_{2(exp)} = 21 \div 23$ MeV, for four excess
neutrons $E_{4(exp)} = 42 \div 45$ MeV, $E_{6(exp)} = 61 \div 63$ MeV and
$E_{8(exp)} = 77 \div 80$ MeV. Therefore, the parameters  in (23) have been obtained
by fitting both the experimental values of $nn$-pair separation energies known [16]
for 27 $nn$-pairs and the values of $E_{n(exp)}$. Eq. (23) has been used here to
calculate the binding energies for the nuclei with $ Z > 10$, although the empirical
values of the excess neutron separation energies are known (24) and could have been
used for the nuclei with  $2 \leq Z \leq 29$.

In Table 2  the calculated values of the nuclear binding energies of some isotopes
with an integer number of $pn$- and $nn$-pairs are presented with their deviations
$\Delta = E_{exp} - E^b$. The values calculated by Weizs\"{a}cker's formula (6) are
presented in Table 2 with their deviations from experimental values
$\Delta_{W}=E_{exp} - E^b_W$ as well.

\begin{center}
{Table 2. Comparison of experimental values of energies and calculated ones. All
energies and deviations are given in MeV.} {\begin{tabular}{ccccccccc|cccccc}\hline
$Z$ & $N_{n}$& $E_{exp}$& $\Delta$& $\Delta_{W}$
&$E_{\alpha}$&$E_{th}$&$E_d$&$E_{th}$&$N_{n}$&$E_{exp}$
&$\Delta$&$N_{n}$&$E_{exp}$&$\Delta$\\\hline
   2 & 0 &    28&  0  & 0 & 28 &28 & 26& 26  & 2 & 29 & 0&  0&28& 0 \\
   3 & 0 &    32&  0  & 9 & 30 &30 &  4& 4   & 2 & 41 & 0&  0&32& 0 \\
   4 & 0 &    56&  0  & 0 & 28&28  & 25& 25  & 2 & 65 & 0& 0& 56& 0 \\
   5 & 0 &    65&  0  & 4 & 33&33  &  8&  8  & 2 & 80 & 0& 0& 65& 0 \\
   6 & 0 &    92&  0  & 2 & 36&37  & 27& 27  & 2 &105 & 0& 0& 92& 0 \\
   7 & 0 &   105&  0  & 7 & 40&36  & 12& 13  & 2 &118 & 0& 0&105& 0 \\
   8 & 0 &   128&  0  & 2 & 35&36  & 23& 22  & 2 &139 & 0& 0&128& 0 \\
   9 & 0 &   137& -2  & 2 & 33&35  & 10& 13  & 2 &154 &-2& 0&137&-2 \\
\hline
\end{tabular}}
\end{center}
\newpage
\begin{center}
{Table 2. Continued} {\begin{tabular}{ccccccccc|cccccc}\hline $Z$&
$N_{n}$&$E_{exp}$& $\Delta$& $\Delta_{W}$
&$E_{\alpha}$&$E_{th}$&$E_d$&$E_{th}$&$N_{n}$&$E_{exp}$
&$\Delta$&$N_{n}$&$E_{exp}$&$\Delta$\\\hline
  10 & 0 &   161& -3  &-1 & 33&36  & 23& 22  & 2 &177 &-3& 0&161&-3 \\
  11 & 0 &   174& -3  & 1 & 37&35  & 14& 14  & 2 &193 &-5& 0&174&-3 \\
  12 & 0 &   198& -1  & 1 & 38&36  & 24& 22  & 2 &216 &-3& 0&198&-1 \\
  13 & 0 &   212& -1  & 2 & 38&35  & 14& 14  & 2 &233 &-1& 0&212&-1  \\
  14 & 0 &   237&  2  & 3 & 38&36  & 25& 22  & 2 &256 & 0& 0&237& 2 \\
  15 & 0 &   251&  2  & 4 & 39&35  & 14& 14  & 2 &271 & 1& 0&251& 2 \\
  16 & 0 &   272&  2  & 2 & 36&36  & 21& 22  & 2 &292 & 1& 0&272& 2 \\
  17 & 0 &   286&  2  & 3 & 35&35  & 14& 14  & 2 &307 & 2& 0&286& 2 \\
  18 & 4&344    & -3   &-3 & 35&36  & 20& 22  & 4 &344 & -3& 0&307& 1 \\
  19 & 2& 342   & 1   &-1 & 35&35  & 14& 14  & 6 &376 & -5& 0&321& 1 \\
  20 & 0& 342   & 1   & 1 & 35&36  & 21& 22  & 8 &415 & -6& 0&342& 1 \\
  21 & 2& 377   & 0   &-2 & 34&35  & 13& 14  & 8 &432 & -3& 0&355& 0 \\
  22 & 4& 419   & 0  &-2 & 38&36  & 22& 22  & 6 &438 &   0& 0&376&-1\\
  23 & 4& 435   & 3  & 0 & 35&35  & 15& 14  & 6 &453 & 1& 0&390& 0 \\
  24 & 4& 456   & 2  & 0 & 38&36  & 22& 22  & 8 &489 &-4& 0&412&-1 \\
  25 & 4& 472   & 4  & 2 & 37&35  & 15& 14  & 6 &489 & 2& 0&427& 0 \\
  26 & 4& 492   & 3  & 1 & 36&36  & 20& 22  & 6 &510 & 1& 0&448& 0 \\
  27 & 4& 507   & 4  & 2 & 36&35  & 15& 14  & 8 &541 & -1& 0&463& 1 \\
  28 & 2& 507   & 2  & 4 & 36&36  & 20& 22  & 8 &562 & -2& 0&484& 1 \\
  29 & 4& 541   & 2  & 1 & 35&35  & 13& 14  & 8 &576 &-1& 0&497& 0 \\
  30 & 4& 559   & 5   & 0 & 32&29  & 19&16 & 8 & 625 & 2& 2&538& 4 \\
  31 & 6& 592   & 5   &-2 & 31&29  & 14&13 &10 & 641 & 1& 2&551& 4 \\
  32 &10& 646   & 0   &-2 & 35&34  & 20&21 &14 & 677 &-4& 6&611& 3 \\
  33 &10& 660   & 2   &-3 & 34&33  & 14&13 &14 & 692 &-2& 6&624& 3 \\
  34 &12& 697   & 0   &-1 & 34&33  & 20&20 &14 & 713 &-1& 6&643& 2 \\
  35 &10& 694   & 3   &-3 & 34&33  & 14&13 &14 & 728 & 1& 6&657& 3 \\
  36 &12& 732   & 3   & 0 & 35&33  & 20&20 &16 & 762 &-2& 8&695& 3 \\
  37 &12& 748   & 6   & 1 & 36&32  & 16&13 &16 & 777 & 0& 8&709& 4 \\
  38 &12& 768   & 7   & 3 & 36&32  & 21&20 &16 & 796 &-1& 8&729& 4 \\
  39 &12& 782   & 8   & 2 & 34&32  & 14&13  &16 & 811& 1& 8&743& 5 \\
  40 &10& 784   & 8   & 3 & 31&32  & 20&20  &16 & 829& 0&10&784& 8 \\
  41 &10& 797   & 8   & 2 & 35&32  & 13&12  &16 & 842& 1& 8&777& 7 \\
  42 &14& 846   & 3   &-1 & 31&32  & 18&19  &16 & 860& 0& 8&797& 7 \\
  43 &12& 844   & 6   & 0 & 32&31  & 14&12  &16 & 875& 2& 8&809& 7 \\
  44 &14& 878   & 3   &-1 & 32&31  & 18&19  &16 & 893& 2& 8&827& 6 \\
  45 &14& 891   & 4   &-2 & 32&31  & 13&12  &16 & 907& 3&10&858& 6 \\
  46 &16& 925   & 3   &-1 & 31&31  & 19&19  &18 & 954& 2&10&876& 6 \\
  47 &14& 923   & 5   &-1 & 32&31  & 13&12  &18 & 983& 2&10&888& 6 \\
  48 &18& 973   & 3   &-1 & 32&31  & 19&18  &20 & 987& 1&12&924& 5 \\
  49 &16& 970   & 5   &-1 & 32&30  & 13&12  &20 &1002& 3&12&936 & 5\\
  50 &20&1021   & 4   & 2 & 33&30  & 19&18  &22 &1050& 4&14&971 & 5\\
  51 &20&1033   & 5   & 1 & 30&30  & 13&12  &22 &1049& 4&16& 998& 4\\
  52 & 26&1096  &  3   & 7  &30 &30  &18 &18  &28 &1110& 3&20&1050& 4\\
  53 & 22&1079  &  5   & 2  &31 &30  &13 &12  &26 &1136& 5&18&1045& 4\\
  54 & 24&1113  &  5   & 4  &31 &30  &18 &18  &26 &1142& 5&18&1063& 4\\
  55 & 24&1126  &  6   & 4  &31 &29  &13 &12  &28 &1155& 5&20&1092& 5\\
  56 & 26&1159  &  7   & 6  &31 &29  &17 &17  &28 &1170& 3&22&1125& 6\\
  57 & 24&1156  &  7   & 4  &31 &29  &13 &12  &26 &1170& 6&20&1121& 5\\
  58 & 24&1173  &  7   & 4  &30 &29  &17 &17  &28 &1209& 2&20&1138& 4\\
\hline
\end{tabular}}
\end{center}
\newpage
\begin{center}
{Table2. Continued}
{\begin{tabular}{ccccccccc|cccccc} \hline $Z$ & $N_{n}$&
$E_{exp}$& $\Delta$&$\Delta_{W}$
&$E^{sep}_{\alpha}$&$E_{th}$&$E^{sep}_d$&$E_{th}$&$N_{n}$&$E_{exp}$&$\Delta$
&$N_{n}$&$E_{exp}$&$\Delta$\\\hline
  59 & 22&1169  &  7   & 2  &30 &29  &13 &12  &26 &1209& 4&22&1166& 7\\
  60 & 22&1185  &  6   & 3  &26 &29  &17 &13  &28 &1238& 1&22&1179& 6\\
  61 & 24&1210  &  4   & 0  &26 &28  &11 &12  &28 &1237& 1&20&1175& 5\\
  62 & 28&1253  &  1   &-1  &28 &28  &17 &17  &32 &1280&-1&24&1224& 3\\
  63 & 26&1251  &  1   &-2  &27 &28  &11 &12  &30 &1280& 1&20&1203& 3\\
  64 & 30&1296  &  1   & 0  &29 &28  &16 &16  &32 &1309& 0&24&1252& 1\\
  65 & 28&1294  &  2   &-1  &29 &28  &13 &11  &30 &1309& 2&22&1247& 0\\
  66 & 32&1338  &  1   & 1  &29 &28  &16 &16  &34 &1351& 1&26&1295& 1\\
  67 & 30&1336  &  2   &-1  &28 &27  &12 &11  &32 &1350& 2&28&1322& 1\\
  68 & 30&1351  &  1   & 0  &27 &28  &15 &16  &34 &1391& 1&26&1322&-1\\
  69 & 32&1378  &  2   & 0  &27 &27  &12 &11  &34 &1392& 2&28&1349& 1\\
  70 & 34&1407  &  2   & 1  &28 &27  &15 &16  &36 &1419& 1&28&1365& 0\\
  71 & 32&1404  &  2   & 0  &26 &27  &12 &11  &36 &1431& 2&28&1376&-1\\
  72 & 36&1446  &  1   & 0  &26 &27  &15 &16  &38 &1459& 1&30&1406&-1\\
  73 & 34&1445  &  2   & 0  &27 &27  &12 &11  &38 &1471& 2&32&1432& 1\\
  74 & 38&1486  &  1   & 1  &27 &27  &15 &15  &40 &1498& 1&32&1447&-1\\
  75 & 36&1484  &  2   & 0  &26 &26  &12 &11  &40 &1510& 2&34&1473& 1\\
  76 & 40&1526  &  3   & 2  &28 &26  &16 &15  &40 &1526& 3&34&1488& 0\\
  77 & 38&1524  &  2   & 1  &26 &26  &11 &11  &40 &1538& 4&34&1499&-1\\
  78 & 38&1540  &  3   & 2  &27 &26  &15 &15  &42 &1567& 6&36&1527& 1\\
  79 & 40&1566  &  6   & 4  &28 &26  &12 &11  &42 &1580& 7&36&1538& 1\\
  80 & 42&1595  &  8   & 7  &28 & 26&15& 15  &44 &1609&10&36&1553& 2\\
  81 & 42&1608  &  9   & 8  &28 & 25&12&  11 &46 &1641&11&38&1577& 4\\
  82 & 44&1636  & 12   & 11 &28 & 25&15&  14 &48 &1663& 7&40&1605& 7\\
  83 & 42&1633  &  9   & 9  &25 & 25&10&  11 &46 &1664& 7&38&1603& 4\\
  84 & 40&1631  &  5   & 7  &23 & 25&13&  14 &44 &1686& 6&38&1617& 2\\
  85 & 40&1641  &  4   & 6  &23 & 25&10&  11 &48 &1685& 2&38&1627& 1\\
  89 & 50&1742  & -2   &-4  &23 & 24&10&  11 &50 &1742&-2&46&1727&-3\\
  90 & 52&1767  & -1   &-4  &24 & 24&14&  14 &54 &1778& 0&48&1752&-3\\
  91 & 50&1765  & -2   &-4  &24 & 24&10&  11 &52 &1777&-1&46&1751&-5\\
  92 & 54&1802  &  0   &-4  &24 & 24&14&  13 &56 &1812& 0&50&1787&-3\\
  93 & 52&1801  & -2   &-3  &24 & 24&11&  11 &54 &1812&-1&48&1786&-5\\
  94 & 58&1847  &  1   &-4  &23 & 24&13&  13 &58 &1847& 1&54&1832&-1\\
  95 & 52&1824  & -3   &-3  &23 & 24&10&  10 &56 &1846& 0&52&1832&-3\\
  96 & 52&1836  & -3   &-3  &22 & 24&12&  13 &54 &1848&-2&48&1823&-7\\
  97 & 52&1846  & -4   &-3  &23 & 23&10&  10 &56 &1869&-1&54&1866&-2\\
  98 & 54&1870  & -3   &-3  &22 & 23&12&  13 &54 &1870&-3&50&1831&10\\
  99 & 52&1874  &  1   & 4  &22 & 23& 9&  10 &56 &1891&-2&50&1879&-4\\
 100 & 54&1891  & -5   &-3  &21 & 23&12&  13 &60 &1930*&5&50&1865&10\\
 101 & 54&1903* & -3   & 0  &   & 23&  &   10&60 &1940*&5&50&1877&-8\\
 102 & 56&1927* & -1   & 0  &   & 23&  &   12&60 &1951*&3&52&1901&-7\\
 103 &56  &1937*&-2 & 0&    &22 &  & 10&62 & 1972*& 5 &52&1910*& -8\\
 104 &58  &1960*&-1 & 0&    &23 &  & 12&62 & 1984*& 5 &54&1935*& -6\\
 105 &58  &1969*&-2 & 0&    &22 &  & 10&62 & 1993*& 4 &54&1943*& -8\\
 106 &60  &1993*& 0 & 0&    &22 &  & 12&64 & 2016*& 6 &56&1967*& -6\\
 107 &60  &2001*&-1 & 0&    &22 &  & 10&64 & 2025*& 5 &56&1976*& -7\\
 108 &60  &2012*&-2 & 0&    &22 &  & 12&64 & 2036*& 4 &56&1986*& -9\\
 109 &64  &2045*& 3 & 0&    &22 &  & 10&68 & 2068*& 9 &60&2021*& -3\\
 110 &64  &2056*& 2 & 0&    &21 &  & 12&68 & 2079*& 8 &60&2031*& -5\\
 111 &64  &2064*& 1 & 0&    &21 &  & 10&68 & 2088*& 7 &60&2039*& -7 \\
\hline
\end{tabular}}
\end{center}

\begin{center}
{Table2. Continued} {\begin{tabular}{ccccccccc|cccccc} \hline $Z$ & $N_{n}$&
$E_{exp}$& $\Delta$&$\Delta_{W}$
&$E^{sep}_{\alpha}$&$E_{th}$&$E^{sep}_d$&$E_{th}$&$N_{n}$&$E_{exp}$&$\Delta$
&$N_{n}$&$E_{exp}$&$\Delta$\\\hline
 112 &68  &2099*& 6 & 0&    &21 &  & 11&70 & 2110*&10 &64&2075*&  0 \\
 113 &70  &2119*& 8 & 0&    &21 &  & 10&72 & 2129*&11 &66&2095*&  1 \\
 114 &68  &2117*& 4 & 0&    &21 &  & 11&70 & 2129*& 8 &64&2093*& -4 \\
 115 &68  &2126*& 2 & 0&    &21 &  & 10&72 & 2148*& 9 &64&2101*& -5 \\
 116 &68  &2136*& 1 & 0&    &21 &  & 11&72 & 2159*& 9 &64&2123*&  4\\
 117 &68  &2143*&-1 & 0&    &21 &  & 10&72 & 2167*& 7 &64  &2118 & -9\\
 118 &68  &2153*&-2 & 0&    &21 &  & 11&72 & 2177*& 6 &64  &2128 &-10\\
 119 &68  &2161*&-4 & 0&    &20 &  &  9&72 & 2185*& 5 &64  &2135 &-13\\
\hline
\end{tabular}}
*The value is calculated by Weizs\"{a}cker's formula (6)
\end{center}
One can see from this Table that the values of $\Delta$ and $\Delta_{W}$ are
comparable. To test the formulas for $E^{nuc}$, $E^{st}$ and $E^{C}$ independently
of the energy of excess neutrons, the energy of the alpha-particle separation
energy, as well as the deutron separation energy, have been calculated. In the Table
the experimental values $E_\alpha$ and $E_d$ are followed by the calculated values
$E_{th}$. The left part of the Table contains values for nuclids of the most
abundant isotopes for even $Z$ or nearest to them for odd $Z_1$. The right part
contains the values of binding energies for the isotopes with large and small number
of $N_{n}$. Since for the nuclei with $Z \leq 10$ the experimental values of the
energy of excess neutrons (24) is used, the values of $\Delta$ in the cases with a
larger number of neutrons are the same as for nuclei with $N = Z$.

\section{Conclusion}

There are two specific assumptions in the alpha-cluster model presented here.
Namely, (1) the $pn$-pair interactions are responsible for building alpha-clusters
and their adherence and (2) the proton and the neutron from a pair have equivalent
bound state nuclear potentials. Formulas to calculate the radii and binding energies
of nuclei have been obtained in the framework of the model. The deviation of the
calculated radii from the experimental values is $<\Delta^2>^{1/2} = 0.05$ Fm, which
is $\sim 1\%$ of the magnitude of radius. The calculated values of binding energies
of the nuclei with an integer number of $pn$-pairs and $nn$-pairs have deviations
from the experimental values comparable with the values due to Weizs\"{a}cker's
formula and they have  less than $ 0.5\%$ on average.

The calculations do not take into account the spin-orbit correlations and
consequently shell effects stay beyond this consideration. Therefore, the accuracy
obtained for the calculated values may provide some evidence on the influence of the
shell effects on the values of the charge radii and binding energies for the nuclei
in their ground states. The accuracy of several hundredths of Fm for the radius and
of several MeV for the binding energy reveal the possible range of the shell effect
influence. It is in an agreement with the range of applicability of shell model in
the various tasks of nuclear spectroscopy.

The obtained values of the Coulomb repulsion and the nuclear force energies of the
cluster-cluster interaction, which are 1.925 MeV and 4.350 MeV correspondingly, can
be useful in a microscopic description of light nuclei $^{12}C$ and $^{16}O$ in the
framework of the few-body task.

The obtained formulas to calculate the Coulomb radius $R_C$, the charge radius $R$
and the radius of the last proton position $R_p$ by the number of $\alpha$-clusters
show that for light nuclei $R < R_C \approx R_p $ and for heavy nuclei $R < R_C <
R_p$.

The empirical values of the Coulomb energy $E^C$ for the nuclei with $N = Z$ have
been obtained. These values together with the empirical values of the short-range
nuclear force energy, $E^{shr} = E_{exp} + E^C$, can be useful in studying the
heavy-ion induced reactions such as fusion or fission reactions.

\section*{Acknowledgments}

I would like to thank Prof. V. B. Belyaev for helpful discussions about the results
of the work.

\end{document}